\documentclass[Physsubmission, Phys]{SciPost}
\usepackage{caption}
\usepackage{subcaption}

\hypersetup{
    colorlinks,
    linkcolor={red!50!black},
    citecolor={blue!50!black},
    urlcolor={blue!80!black}
}

\usepackage[bitstream-charter]{mathdesign}
\DeclareSymbolFont{usualmathcal}{OMS}{cmsy}{m}{n}
\DeclareSymbolFontAlphabet{\mathcal}{usualmathcal}

\begin{document}

\begin{center}{\Large \textbf{
Intermittency in $pp$ collisions at $\sqrt{s}=$ 0.9, 7 and 8 TeV from the CMS collaboration
}}\end{center}

\begin{center}
Z. Ong\textsuperscript{$\star$},
P. Agarwal,
H.W. Ang,
A.H. Chan,
C.H. Oh
\end{center}

\begin{center}
Department of Physics, National University of Singapore

* ongzongjin@u.nus.edu
\end{center}

\begin{center}
\today
\end{center}


\definecolor{palegray}{gray}{0.95}
\begin{center}
\colorbox{palegray}{
  \begin{tabular}{rr}
  \begin{minipage}{0.1\textwidth}
    \includegraphics[width=30mm]{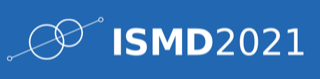}
  \end{minipage}
  &
  \begin{minipage}{0.75\textwidth}
    \begin{center}
    {\it 50th International Symposium on Multiparticle Dynamics}\\ {\it (ISMD2021)}\\
    {\it 12-16 July 2021} \\
    \doi{10.21468/SciPostPhysProc.?}\\
    \end{center}
  \end{minipage}
\end{tabular}
}
\end{center}

\section*{Abstract}
{\bf
The intermittency-type fluctuations as outlined by Bialas and Peschanski in the 1980s is analysed in $pp$ collisions at $\sqrt{s}=$ 0.9, 7 and 8 TeV from the CMS collaboration at CERN. Our preliminary analysis shows that the intermittency exponents in the bin-averaged scaled factorial moments decrease in magnitude with increasing collision energy at the TeV scale, which suggests that the cascading nature of multiparticle production described by the $\alpha$-model is weakening. We outline possible areas planned for future studies.
}


\section{Introduction}

Intermittency in multiparticle production, where fluctuations in particle multiplicity can be observed at all scales, can be quantified via the bin-averaged scaled factorial moments \cite{Bialas:1985jb, Bialas:1988wc}:
\begin{equation}
	\label{eq:FactorialMoments}
	F_{q}(\delta \eta) = M^{q-1} \sum_{m=1}^{M} \frac{\langle n_{m} (n_{m}-1) \ldots (n_{m}-q+1) \rangle}{\langle N \rangle ^{q}}
\end{equation}
where $F_{q}$ is the $q^{\text{th}}$ bin-averaged moment, $M$ is the number of bins that the (pseudo)rapidity space is divided into (each with size $\delta \eta$), $n_{m}$ is the multiplicity in bin $m$, $N$ is the total event multiplicity and $\langle \ldots \rangle$ represents an average over events. Equation (\ref{eq:FactorialMoments}) is sometimes known as the ``horizontal moments'' in intermittency literature.

In Bialas and Peschanski's formulation \cite{Bialas:1985jb, Bialas:1988wc}, a system is said to be intermittent if a power-law relation $F_q \sim (\delta \eta)^{-\phi_q}$ exists. We expect an intermittent system to produce a straight line in a $\ln{F_q}$ vs. $\ln{(1/\delta \eta)}$ plot towards $\delta \eta \rightarrow 0$, with some positive gradient $\phi_q$ called the ``slope parameter'' or ``intermittency exponent''. The magnitude of $\phi_q$ reflects the strength of intermittency of the system.

\section{Motivation and Objectives}

With the LHC at CERN ushering in a new era of TeV-scale high energy physics, it is worthwhile to reinvestigate this phenomenon in a new energy regime. In our previous work \cite{Ong:2019dlz}, we found that intermittent phenomenon exists in $pp$ collisions at $\sqrt{s} = $ 7 TeV. That motivates our current investigation, where the analysis is extended to include the other available energies in Run 1 from the CERN Open Data Portal -- 0.9 and 8 TeV from the CMS collaboration.

Hence we aim to:
(1) compute the bin-averaged scaled factorial moments and produce log-log plots of $F_q$ vs. $(1/\delta \eta)$ for \textit{pp} collisions at $\sqrt{s}=$ 0.9, 7 and 8 TeV data from the CERN Open Data Portal, and
(2) obtain the intermittency exponents for \textit{q} = 2, 3, 4 and 5 from the data and make observations about any trends that may appear.

%

\section{Data Processing}

MinimumBias datasets from the CERN Open Data Portal are used for this analysis -- 0.9 TeV data from the Commissioning run of 2010~\cite{900GeV}, 7 TeV data from RunA of 2010~\cite{7TeV} and 8 TeV data from RunB of 2012~\cite{8TeV}. About one million events for each energy were included in this analysis.

\subsection*{Event Selection}

Vertices are required to have number of degrees of freedom greater than 4, and be within 15 cm of the beamspot in the \textit{z}-direction. If more than one primary vertex is present, the one with the highest number of associated tracks is chosen to be the primary vertex.

To remove secondaries from background and pileup, tracks are required to have impact parameter significances $d_{0} / \sigma_{d_0}$ and $d_{z} / \sigma_{d_z}$ to be smaller than 5. The relative uncertainty in the momentum measurement $\sigma_{p_\text{T}} / p_{\text{T}}$ is required to be less than 10\%. To ensure good reconstruction efficiency, only \texttt{highPurity} tracks with $p_{\text{T}} >$ 0.5 GeV/c are selected.

\section{Results}

\begin{figure}[h!]
	\centering
	\begin{subfigure}[b]{0.49\linewidth}
		\includegraphics[width=\textwidth]{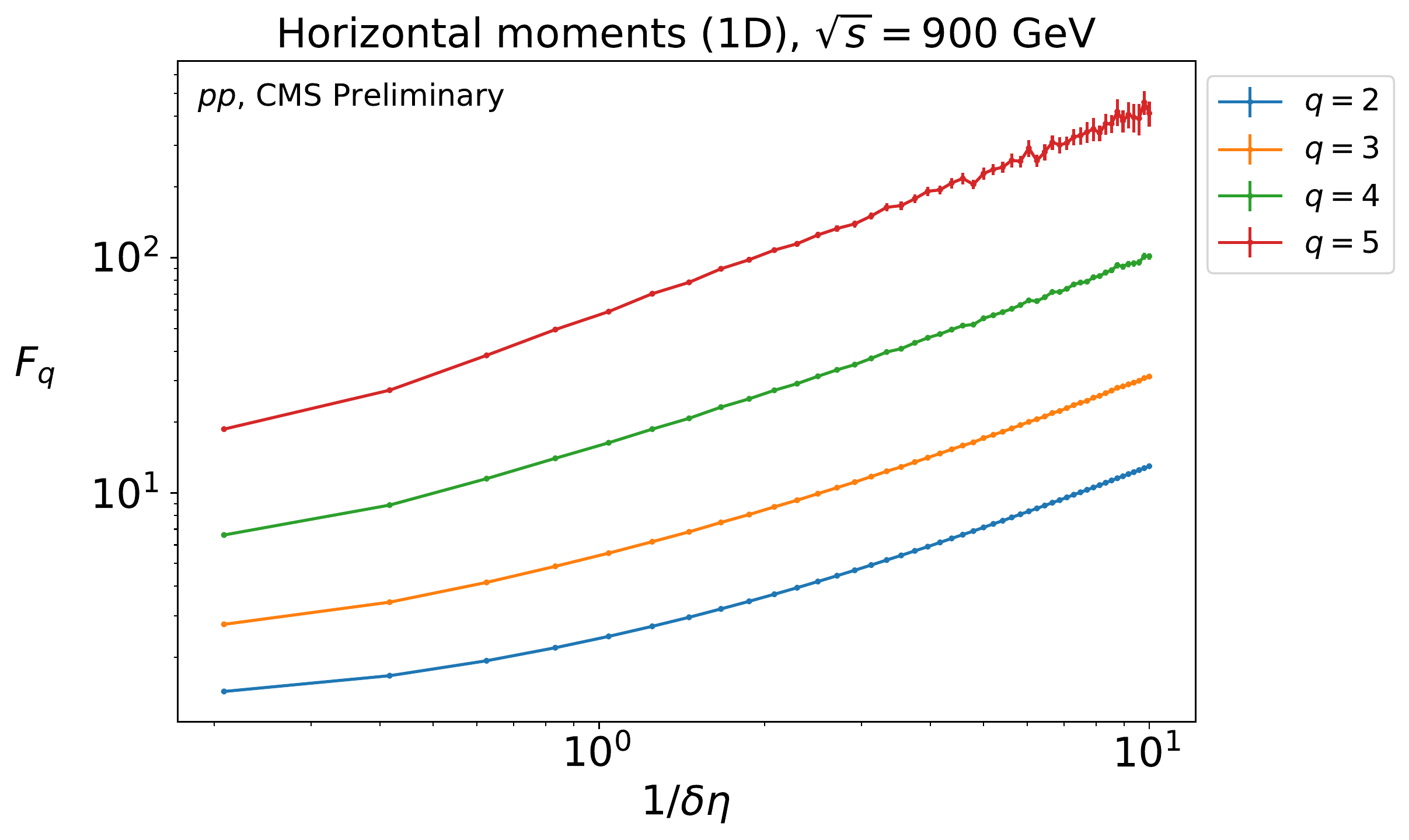}
		\label{fig:900GeV}
	\end{subfigure}
	\begin{subfigure}[b]{0.49\linewidth}
		\includegraphics[width=\textwidth]{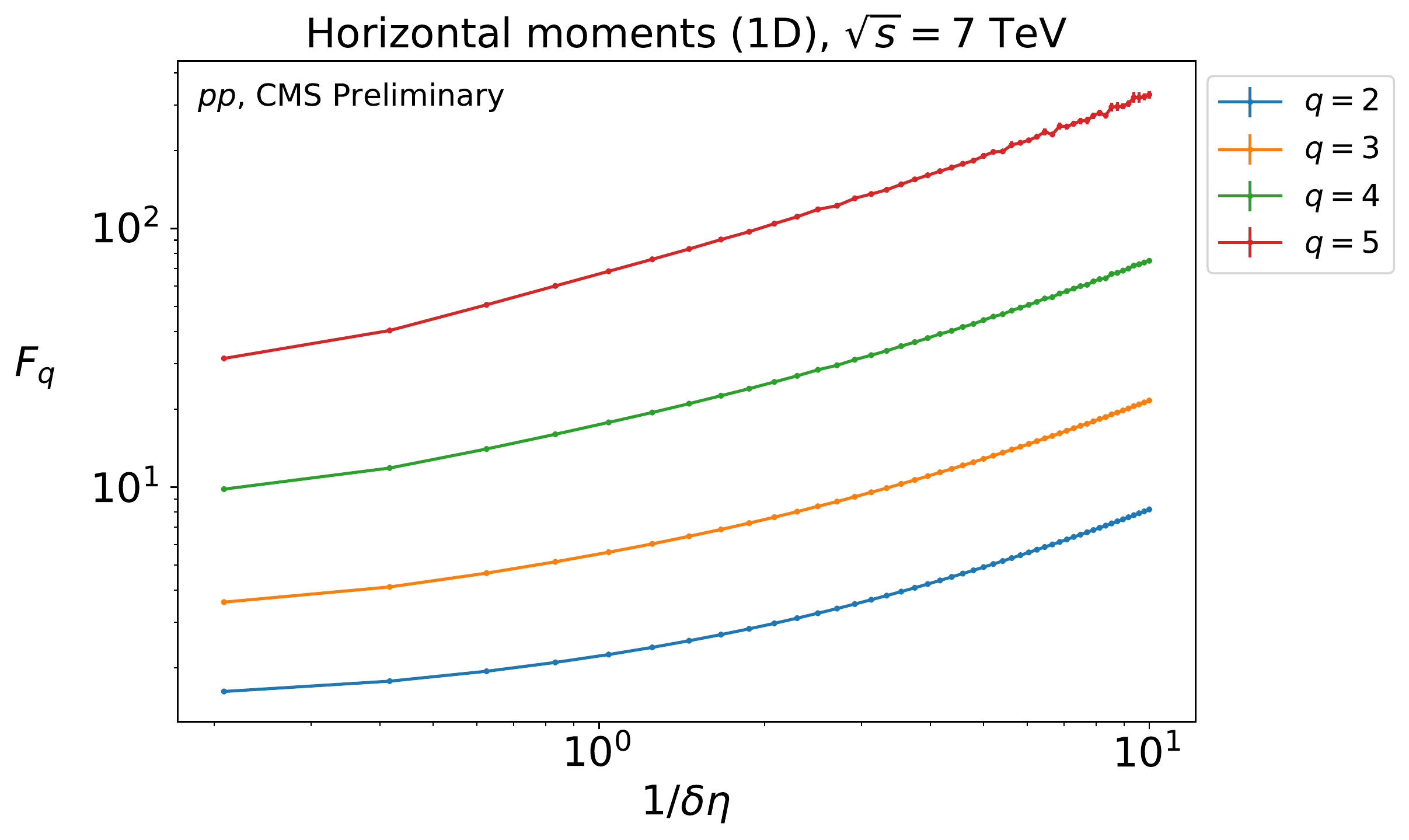}
		\label{fig:7TeV}
	\end{subfigure}
	\begin{subfigure}[b]{0.49\linewidth}
		\includegraphics[width=\textwidth]{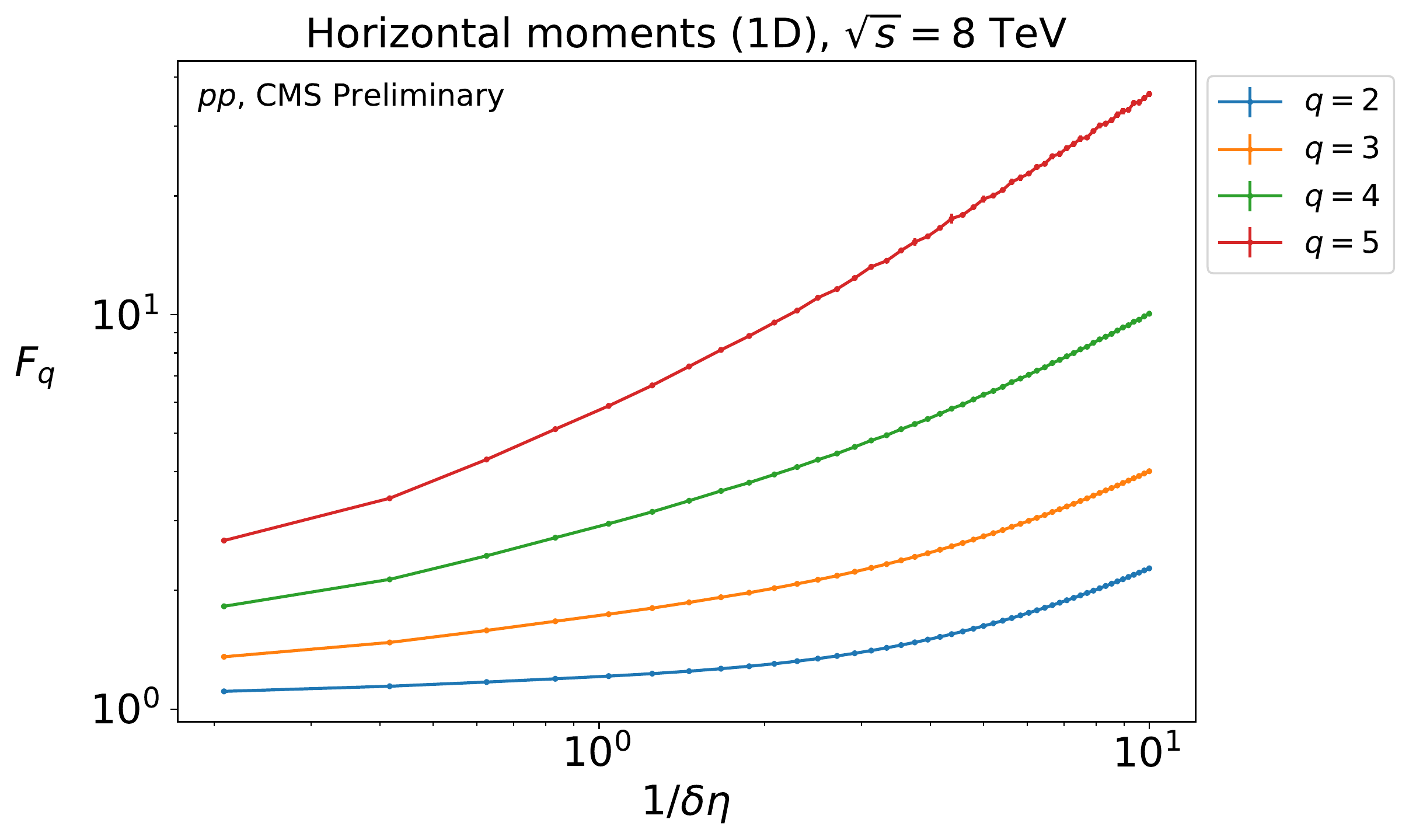}
		\label{fig:8TeV}
	\end{subfigure}
	\caption{Horizontal moments for $\sqrt{s}=$ 0.9, 7 and 8 TeV}
	\label{fig:intermittencyPlots}
\end{figure}

Figure \ref{fig:intermittencyPlots} show the log-log plots in $F_q$ vs $(1/\delta \eta)$ for $\sqrt{s}=$ 0.9, 7 and 8 TeV with error bars plotted (only statistical; most are too small to be seen). Table~\ref{table:intermittencyExponents} summarises the intermittency exponents, which are obtained via linear regression on the rightmost 10 data points of each set of data\footnote{The results in this section have been updated from what was presented at ISMD2021.}.

\begin{table}[h!]
	\centering
	\begin{tabular}{l l l l}
		\hline
		\textbf{\textbf{\textit{q}}} & \textbf{0.9 TeV} & \textbf{7 TeV} & \textbf{8 TeV} \\
		\hline
		2 & $0.889 \pm 0.003$	& $0.787 \pm 0.002$	& $0.560 \pm 0.003$\\
		3 & $0.89 \pm 0.02$	& $0.795 \pm 0.007$	& $0.612 \pm 0.004$\\
		4 & $0.87 \pm 0.08$		& $0.82 \pm 0.03$	& $0.72 \pm 0.01$\\
		5 & $0.9 \pm 0.3$		& $0.87 \pm 0.09$	& $0.90 \pm 0.03$\\
		\hline		
	\end{tabular}
	\caption{Summary of intermittency exponents}
	\label{table:intermittencyExponents}
\end{table}

\section{Conclusion and Discussion}

From the non-zero values in Table \ref{table:intermittencyExponents}, we can immediately conclude that intermittency still persists in $pp$ collisions up to $\sqrt{s} = $ 8 TeV. 

With the exception of $q=5$, the intermittency exponents generally appear to show a decreasing trend as $\sqrt{s} =$ increases from 0.9 TeV to 8 TeV. For $q=5$, the intermittency exponents appear to be equal within errors. These findings are made with data treatment kept uniform across all three energies.

\subsection*{Scaling, Universality and Branching}

The results clearly show that the scaling relation $F_q \sim (\delta \eta)^{-\phi_q}$ still holds at the energies being investigated. However, universality, as noted by Bialas~\cite{Bialas:1992vh} where the intermittency exponents are the same in all experiments, does not seem to hold. More analysis needs to be done before further conclusions can be made about this.

The concept of intermittency is closely related to parton branching processes in multiparticle production~\cite{Bialas:1992vh}. The decreasing intermittency exponents lead us to surmise that as the centre-of-mass energy increases into the TeV scale, other competing processes in multiparticle production might become increasingly important. Still, parton branching ideas remain valid and are a present area of study (see for example,~\cite{Ong:2020vpl}).

\section{Forthcoming Research}

This analysis can be further extended in the following areas:

\begin{enumerate}
	\item \textbf{Unfolding (corrections)}: In this analysis, the particle multiplicities are derived by counting charged tracks from the tracker in CMS. The multiplicity distributions need to be unfolded back to the charged hadron level, so that we can study the intermittency effects from a QCD perspective. Unfolding will also correct for detector-related inefficiencies.
	
	\item \textbf{Higher energies}: With the expected release of Run 2 data to the public, we can look forward to extending this study up to $\sqrt{s} = $ 13 TeV, which will shed more light into intermittent behaviour at the TeV scale.
	
\end{enumerate}

\section*{Acknowledgements}
The authors would like to thank the National University of Singapore, and the support and helpful discussions with colleagues. We would also like to thank the CMS collaboration at CERN for the well-organised release of collision data to the public on the CERN Open Data Portal. This work is also supported by the NUS Research Scholarship.

\bibliography{References.bib}

\nolinenumbers

\end{document}